# Condition for Steep Density Gradient at Separatrix

T. J. Dolan, NPRE Department, University of Illinois, Urbana, IL, 61801 USA

**Abstract**

The slope of the plasma density gradient at the separatrix is influenced by the curvature of the density profile in the scrape-off layer (SOL). This curvature is determined by the parallel flow loss rate and the ionization rate in the SOL, which are influenced by the neutral gas density and electron heating in that region. The radial electric field is also influenced by the ion pressure gradient just inside the separatrix.

**Density Curvature in the SOL**

Enhanced confinement is sometimes observed when the plasma edge density and temperature gradients become very steep. Plasma conditions in the SOL can affect the boundary condition at the separatrix. In the slab approximation a necessary condition for steep density gradients near the separatrix is that the density profile curvature in the SOL be concave upwards ($d^2n/dx^2 > 0$), because both n and dn/dx must be continuous at the separatrix. This is illustrated in Figure 1. When $d^2n/dx^2 < 0$ it is difficult for the density gradient just inside the separatrix to be very steep.

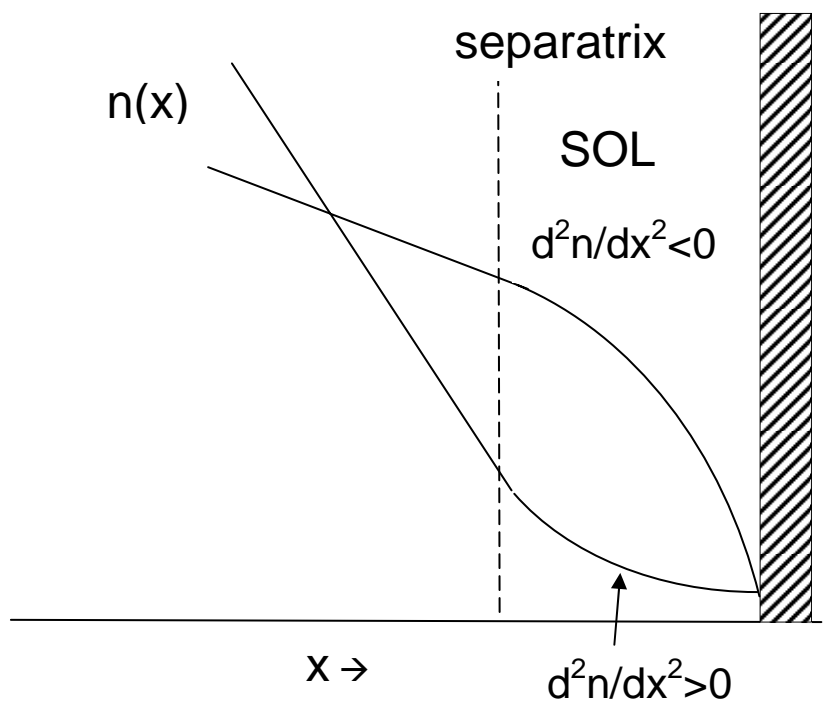

Figure 1. Effect of curvature in the SOL on edge density gradient.

Ignoring recombination, a simple conservation equation for plasma density in the SOL may be written as

$$\partial n/\partial t - (\partial D/\partial x)(\partial n/\partial x) - D\partial^2 n/\partial x^2 \approx \Sigma\, nn_j \langle\sigma v\rangle_j - n/\tau_p \qquad (1)$$

where D is the diffusion coefficient, $n_j$ is the density of neutral species j, $\langle\sigma v\rangle_j$ is the corresponding ionization rate parameter, the summation is over the neutral species, $\tau_p$ is the characteristic time for plasma parallel loss by flow along the magnetic field. It is often assumed that $\tau_p \sim L/c_s$, where $c_s$ is the ion sound speed and L is the characteristic connection length to the limiter or target plate. If significant magnetic mirror reflection or electrostatic potential barriers are present, however, the $c_s$ model is inaccurate. Then $\tau_p$ can be estimated in terms of the mirror ratio, ion-ion collision time, and potential barrier height.[1,2]

At equilibrium the density profile curvature is

$$d^2n/dx^2 = (1/\tau_p - \Sigma\, n_j \langle\sigma v\rangle_j)\, n/D - (1/D)(dD/dx)(dn/dx)] \qquad (2)$$

Often $\partial D/\partial x > 0$ and $\partial n/\partial x < 0$, making the last term positive. The curvature is influenced by the neutral concentrations and by the electron temperature, which affects the ionization rate parameter and the sound speed. With much neutral gas present,

$$\Sigma\, n_j \langle\sigma v\rangle_j > 1/\tau_p\ . \qquad (3)$$

Unless the dD/dx term is large, it is likely that $d^2n/dx^2 < 0$, and the curvature is concave downwards, preventing a steep gradient near the separatrix. If gas recycling from the wall is reduced, then the ionization source term may be reduced enough that

$$\Sigma\, n_j \langle\sigma v\rangle_j < 1/\tau_p\ . \qquad (4)$$

Now the curvature $d^2n/dx^2 > 0$ is concave upwards, and the density gradient near the separatrix can become steep. Of course, in a torus these parameters vary with toroidal and poloidal location. This analysis simply illustrates the idea that SOL conditions may influence the density gradient inside the separatrix.

**Control**

To produce a steep edge density gradient a core source of ions is also required. The density curvature transition in the SOL from concave downwards to concave upwards could be promoted by the following conditions:

- Reduction of recycling and gas emission from chamber walls by wall conditioning, such as by bakeout, boronization or lithium coatings, and discharge cleaning.

- Reduction of gas puffing, which deposits neutrals in the SOL. Pellet injection could deposit less gas in the SOL.

- Heating to raise $T_e$ in the SOL. The higher values of $<\sigma v>_j$ may ionize neutral gas close to the wall, lowering the neutral density in the rest of the SOL. There could be a layer near the wall where $d^2n/dx^2 < 0$, but with $d^2n/dx^2 > 0$ over the rest of the SOL, as illustrated in Figure 2. The heating could also increase the parallel flow loss rate.

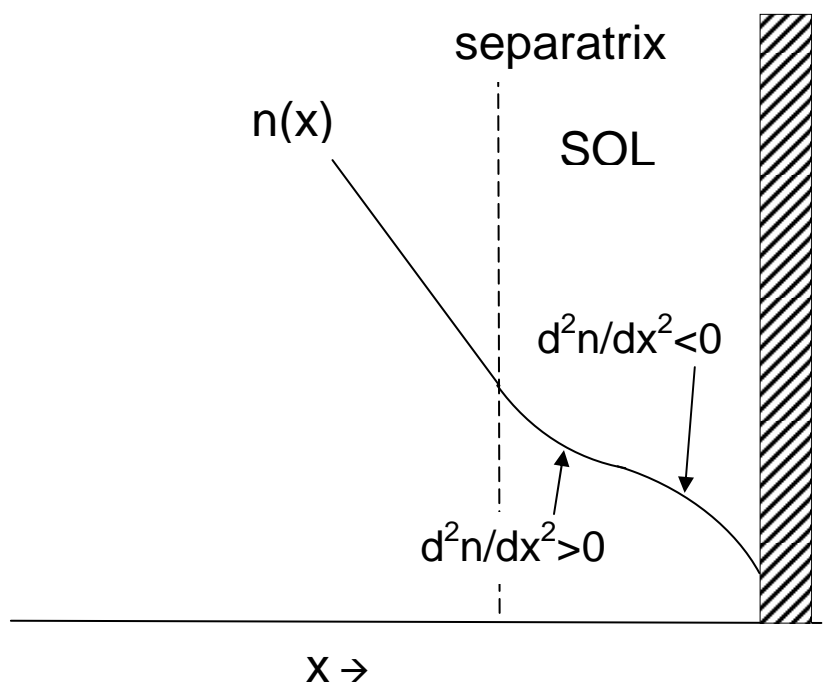

Figure 2. Curvature variation in the SOL. The parallel flow loss term dominates near the separatrix, and the ionization source term dominates near the wall.

The radial electric field just inside the separatrix may be estimated from the equation

$$E_r = (Zen_z)^{-1} dp_z/dr - v_\theta B_\phi + v_\phi B_\theta \quad , \tag{5}$$

where $n_z$ and $p_z$ are the density and pressure of ions with charge Z, e is the electronic charge, $v_\phi$ and $v_\theta$ are the toroidal and poloidal flow velocities, and $B_\phi$ and $B_\theta$ are the toroidal and poloidal magnetic fields.[3] Rapid turbulent electron transport tends to make $E_r$ positive. When the density gradient inside the separatrix becomes steep, however, the large negative value of $dp_z/dr$ can make $E_r$ negative, as has been observed in some transitions from L mode to H mode.

In summary, a concave-upwards density profile in the SOL near the separatrix appears to be necessary for achievement of a steep density gradient inside the separatrix, which is associated with improved plasma confinement. This condition can be promoted by control of the neutral gas density and electron temperature in the SOL.

**Acknowledgment**

The author is grateful to Leonid Zakharov, Graham T. O'Connor and Sizheng Zhu for helpful discussions.

**References.**

1. Eric E. Granstedt and Leonid Zakharov, "Simplest SOL Model", in preparation.

2. V. P. Pastukhov, in Reviews of Plasma Physics, edited by B. B. Kadomtsev (Consultants Bureau, New York, 1984), vol. 13, pp. 203-259.

3. John Wesson, Tokamaks, Third Edition, Clarendon Press, Oxford, UK, 2004. Section 12.5.

_______